# RDAnet: A Deep Learning Based Approach for Synthetic Aperture Radar Image Formation

Andrew Rittenbach and John Paul Walters

University of Southern California, Information Sciences Institute, Arlington, VA 22203

**Abstract**

*Synthetic Aperture Radar (SAR) imaging systems operate by emitting radar signals from a moving object, such as a satellite, towards the target of interest. Reflected radar echoes are received and later used by image formation algorithms to form a SAR image. There is great interest in using SAR images in computer vision tasks such as classification or automatic target recognition. Today, however, SAR applications consist of multiple operations: image formation followed by image processing. In this work, we train a deep neural network that performs both the image formation and image processing tasks, integrating the SAR processing pipeline. Results show that our integrated pipeline can output accurately classified SAR imagery with image quality comparable to those formed using a traditional algorithm. We believe that this work is the first demonstration of an integrated neural network based SAR processing pipeline using real data.*

## 1. Introduction

Synthetic Aperture Radar (SAR) [1], [2] is a remote sensing imaging technique used to form images from radar signals. During imaging, a moving device such as a satellite or an unmanned aerial vehicle [3] emits high frequency radar signals in the direction of a region of interest to image. These signals reflect off the target region and are received by an antenna on the same imaging device. After a sufficient number of signal echoes have been received, they are used as input to an image formation algorithm to produce an image. Image contrast is determined by variations in reflectivity of the region being imaged. SAR imaging has many applications including mapping [4], change detection [5], automatic target recognition [6], and environmental monitoring [7].

Currently, SAR processing is a multistep pipeline. First, image formation and correction algorithms transform measured radar echo data to human understandable images. Then, application specific algorithms are applied to focused SAR images for further processing and to extract desired information from the focused image. Over time, there have been many different algorithms developed to form SAR images from echo data. These algorithms use several different approaches such as backprojection [8], compressed sensing [9], or signal processing [10]. Although there has been interest [11], [12], in the development of deep learning based SAR image formation, to our knowledge as of yet there has not been a demonstration of a fully functional approach. Instead, much deep learning based SAR work has been conducted in the fields of classification [13] and target segmentation [14] [15]. In these works, however, the neural networks operate on images that have already been formed using other SAR image formation approaches

In this work, we present an alternative approach to the traditional SAR processing pipeline. In our approach, as shown in Figure 1, all steps of the SAR processing pipeline are integrated into a deep neural network. In the example shown here, input to the network is raw radar data, while network output is focused SAR imagery that has been accurately classified. With our approach, all computational operations within the pipeline are performed through the forward propagation of a single neural network. There is no need to pass processed data from one processing component to another, greatly simplifying the SAR processing pipeline. Beyond simplification of the SAR processing pipeline, however, there are many other reasons why a deep learning based approach to the SAR processing pipeline would be beneficial. For instance, there may be imaging configurations where simultaneous image formation and target classification may result in improved target classification rate [11] when compared to the case when image formation and target classification are performed sequentially.

There are also potential computational advantages in using a deep learning based approach for SAR processing. Traditional SAR formation algorithms make use of both well-established data transforms such as the FFT as well as SAR specific data corrections. To achieve high throughput SAR image formation, all components of the image formation algorithm must be efficiently implemented. In comparison, a deep learning based approach can largely be implemented through matrix multiplication. Therefore, rather than needing to optimize many different components of the image formation algorithm, optimization can be focused on a single key function. Furthermore, a deep learning based implementation can take advantage of the many processing gains that have been developed by the deep learning community, as well as any gains developed in the future, through the use of commonly used deep learning platforms such as TensorFlow [16] or PyTorch [17]. In addition, a deep learning based SAR pipeline also enables use of hardware accelerators developed specifically for tensor operations such as Google's Tensor Processing Units [18].

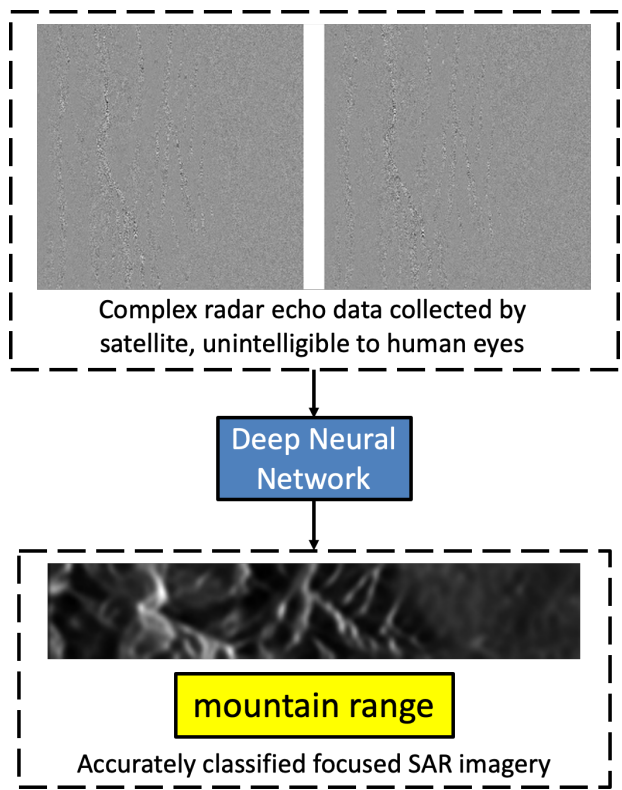

Figure 1: In this work we aim to greatly simplify the SAR processing pipeline using trained neural networks. An example of an integrated pipeline is shown above. Here, the input to the neural network is raw radar data, while the output is a focused SAR image as well as classified imagery.

Finally, a neural network based approach for SAR image formation has the potential to be very robust to both environmental and adversarial factors that can impact the performance of a SAR imaging system such as variations in environmental clutter, atmospheric effects, or adversarial jamming. This is because, with our integrated approach, it is possible to add network layers trained to mitigate the impact of these factors on SAR image quality directly into the image formation and application network.

With that all said, one key factor that is currently keeping the SAR processing pipeline from being integrated within a deep neural network is the lack of suitable neural networks for the SAR image formation problem. We believe that this is due in part to the challenges in obtaining sufficient SAR training data, but largely due to the challenges in identification of an appropriate neural network architecture.

To this end, as part of this work, we present RDAnet, a neural network trained to focus SAR images from raw radar data. This network has been trained to match performance of the Range Doppler Algorithm [10], a well-known signal processing based SAR image formation algorithm. As will be discussed, we trained RDAnet to generate the output of the complete set of operations of the Range Doppler Algorithm by treating the SAR formation problem as a supervised learning problem. We built a dataset containing pairs of raw echo data collected by a research satellite and the echo's corresponding focused image. These echo/image pairs were then used to train RDAnet. After training, the RDAnet was then integrated with a classification network used to identify landscape types in order to demonstrate an integrated SAR processing pipeline.

Taken as a whole, the major contributions of this paper are:
1. We demonstrate, through training and evaluation of RDAnet, the first deep learning based approach to SAR image formation.
2. We demonstrate a proof-of-concept integrated SAR processing pipeline where both image formation and image application processing are performed using a single deep neural network.

The remainder of this paper is laid out as follows: Section 2 details the methods used to build our dataset, the RDAnet architecture, RDAnet training details, and classification network training details. Section 3 shows key results, including comparison of images produced with RDAnet and a traditional approach, and evaluation of the integrated network. Section 4 discusses these results and their implications, Section 5 discusses potential future work, and Section 6 concludes the paper.

## 2. Methods

In this section, we discuss the approach used to collect and preprocess SAR data used for training, the RDAnet architecture, the steps taken to train the RDAnet, and the steps taken to train a classifier used for demonstration of an integrated SAR processing pipeline. The two key components of the RDAnet architecture are shown in Figure 2: a Deep Convolutional Encoder (DCE) followed by a Deep Residual Network (DRN). At a high level, the purpose of the DCE is generate a mapping from raw SAR echo data to a first pass at a focused SAR image. However, we found that SAR images formed by the DCE had poor image resolution and contrast, compared with SAR images formed by traditional approaches. For this reason, the output of the DCE is passed directly to the DRN. The DRN then uses single image super-resolution techniques [20] to upsample the image and improve focused image quality in terms of image resolution and image contrast. Together, the end result is a model that generates focused SAR images from raw echo data.

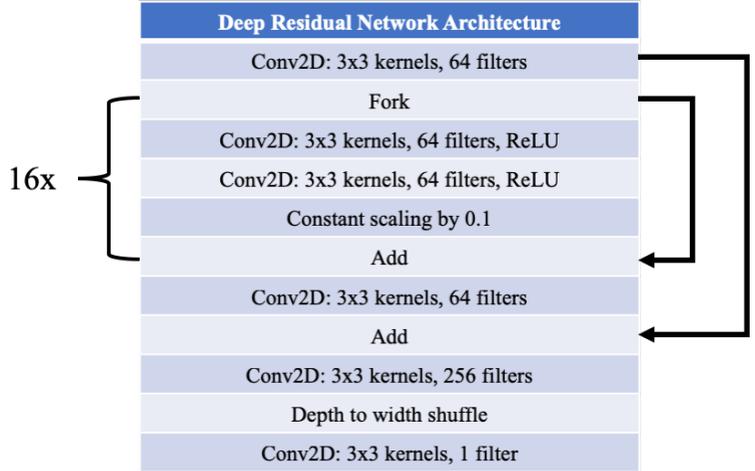

Figure 2: Architecture for the RDAnet Deep Convolutional Encoder (left) and Deep Residual Network (right). Kernel size, number of filters, and activation function for each layer are shown. The arrows show layers that are connected directly to each other and skip intermediate layers.

As mentioned, in this work we made use of the we used the well known Range Doppler Algorithm to focus SAR. Briefly, this algorithm performs the following set of operations to generate a focused SAR image: During processing, echo data is stored in a 2D complex matrix. After the matrix is formed, a matched filter is applied to each matrix row. Once filtering is complete, a 1D FFT is taken along column of the matrix, followed by interpolation along each row of the transformed data. Finally, a second set of matched filters are applied to each column of the matrix. After this, the matrix is transformed back to the spatial domain by an inverse FFT. For some tasks, such as target classification considered in this work, the complex magnitude of each pixel is calculated. In some cases, an additional processing step called 'multilooking' is also performed [19]. There are different approaches to how this step is handled, but the purpose of multilooking is to reduce image speckle. Regardless, the end result is a focused SAR image that can be used for visualization or further processing. In our case, the images were used to train RDAnet.

## 2.1. Data collection and preprocessing

Raw SAR echo data was obtained from the Alaska Satellite Facility (ASF) [21], an organization that allows researchers to access both raw and processed remote sensing data, including SAR data, from a set of satellites. For this work, data collected from the European Remote Sensing (ERS) [22], [23] satellite was used. The ERS-2 was a research satellite launched by the European Space Agency in 1995 and collected data through 2011. The ASF Vertex tool [24] was used to download frames of raw SAR data collected by the ERS-2. In total, 4451 frames were downloaded, containing raw data collected by the ERS-2 while it was orbiting over Alaska. The ERS-2 collected mapping data, therefore, SAR images used in this work consist of mountain landscapes.

To build a training dataset, each radar frame was divided up into unique 4096x4096 segments, resulting in a total of 35641 segments. Each segment was downsampled using MATLAB's resample function [25] to size 256x256. This was done to enable rapid experimentation of different neural network architectures with a dedicated GPU server, to demonstrate the feasibility of generation of SAR images through a deep learning approach. After downsampling the raw data segment, a SAR image was formed using the Range Doppler Algorithm. Finally, to reduce noise, each reconstructed image was smoothed using a gaussian kernel with sigma of 0.75. After all images were formed, SAR image pixel values were scaled such that all pixel values ranged linearly between 0 and 1 throughout the entire dataset.

SAR raw radar data is complex, containing both real and imaginary components. Representative raw echo data is seen in Figure 1. As seen, before processing, the data is nearly unintelligible from the perspective of a person. It is very difficult to see any sort of feature or pattern within the data, highlighting the need for approaches to focus SAR images. For training, all raw data was normalized by the highest complex magnitude in the dataset, bounding the data between -1 and 1. Finally, the raw data was zero centered by calculating the mean of each pixel across the entire echo dataset and then subtracting each pixel in each echo by its respective mean. Raw data was then stored in 256x256x2 arrays, where the real component of the raw was stored in the first channel while the imaginary component was stored in the second channel. After all processing was complete, 33641 echo/image pairs were used for model training, while the remaining 2000 echo/image pairs were used for model validation.

### 2.2. Deep Convolutional Encoder architecture

To map the raw SAR echo data to the SAR image, a convolutional encoder similar to VGG11 [26] was used. The architecture of the convolutional encoder is shown in Figure 2. As shown, the encoder consists of four sets of convolutional blocks with increasing number of filters, from 64 up to 512, each with kernel dimensions 3x3. Every convolutional layer is connected to the next via a leaky rectified linear unit function [27]. Following each convolutional block, two-dimensional max-pooling with stride 2 was used, to reduce the dimensionality of the encoded data by two in each dimension. After the final convolutional block, two-dimensional spatial dropout [28] is used to provide regularization and prevent overfitting. Unlike standard dropout [29], where each unit is set to 0 with a chosen probability at each training iteration, with spatial dropout entire convolutional feature maps are set to 0 instead. For this work, feature maps were set to 0 with probability 0.5. The spatial dropout layer is then connected to two fully connected layers, with 2014 hidden units each. The first fully connected layer is connected to the second via a leaky rectified linear unit function. Finally, the output of the second fully connected layer is reshaped such that the network output is 19x106, which is half the size of the focused SAR image in each dimension. The output of the Deep Convolutional Encoder is then passed to the Deep Residual Network for further processing.

### 2.3. Deep Residual Network architecture

To improve image resolution, an approach inspired by the EDSR single image super resolution technique [30], was used. The network architecture for the Deep Residual Network, also shown in Figure 2, contained 16 residual encoder [31] blocks. As shown, each residual block contains two convolutional layers, with the first connected to the second through a rectified linear unit function. The output of the second convolutional layer is then added to a scaled copy of the input to the first convolutional layer. Each convolutional layer contains 64 filters. Following the last residual block, there is a convolutional layer with 256 filters. As in the DCE, here again each filter uses a 3x3 convolutional kernel. The output of this last convolutional layer is then reshaped using subpixel convolution [32] such that the image size is upsampled by a factor of two in each dimension. After this, the final layer contains a convolution with a single 3x3 filter.

### 2.4. RDAnet training details

The RDAnet was implemented using TensorFlow [16] and trained using the Adam optimizer [33] to minimize the mean absolute error between the output of the network and SAR images reconstructed using the Range Doppler Algorithm. The initial optimizer learning rate was set to 1e-4, all other parameters were left as default. Training minibatch size was 32. Figure 3 shows the mean absolute error of both the training and validation set plotted over each training epoch. As seen, the mean absolute error of the validation set tracks closely with the mean absolute error of the training set throughout the training process, suggesting that the RDAnet generalizes well and that training was stable.

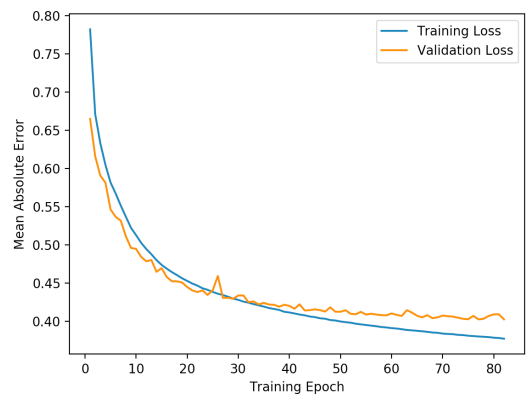

Figure 3: Training and validation loss reported at each training epoch. Mean absolute error was used as the loss function to minimize.

### 2.5. Demonstration Application Network

To demonstrate proof-of-concept of an integrated SAR pipeline, a small classification network was trained using a subset of the data used to train RDAnet. The purpose of this network was to classify two types of landscape imagery:

mountain landscapes and landscapes containing rivers. The architecture of the classifier is shown in Figure 4. As seen, the classifier network consists of two convolutional network layers with 5x5 convolutional kernels, with stride of two in each dimension. The first layer has 64 filters while the second has 128. Both layers are activated with the Leaky ReLU activation function. Furthermore, to prevent overfitting, each convolutional layer is followed by a standard dropout layer with dropout rate of 0.3. After the second dropout layer, the layer is flattened and then passed to a fully connected layer with two hidden units. This layer is then activated with the softmax function, as is standard for classification problems.

The dataset used to train the network consisted of an augmented dataset containing 2000 total training images. While going through our dataset used to train RDAnet, we identified 250 images containing mountain landscapes, and 250 images containing river landscapes. Each of these images were then flipped horizontally, vertically, and both horizontally and vertically in order to expand the number of images to 1000 of each landscape class. The classifier network was then trained using the Adam optimizer with learning rate 1e-4 to minimize sparse categorical crossentropy over 10 epochs with minibatch size of 32.

| Classifier Network Architecture |
| --- |
| Conv2D: 5x5 kernels, 64 filters, Leaky ReLU, stride 2 |
| Dropout |
| Conv2D: 5x5 kernels, 128 filters, Leaky ReLU, stride 2 |
| Dropout |
| Fully Connected Layer, 2 hidden units, softmax |

Figure 4: Architecture for classifier network used to identify different classes of landscape imagery and to demonstrate integrated SAR processing pipeline.

## 3. Results

### 3.1. Comparison to RDA SAR images

SAR images formed using the Range Doppler Algorithm were compared with images generated by the RDAnet. Representative results are shown in Figure 5, which contains side by side comparison between images produced from the same echo data using the Range Doppler Algorithm and RDAnet. As is demonstrated, our model produces SAR images of comparable image quality. The same landscape features are seen in both images, indicating that the raw echo data is being properly focused after passing through the network.

Interestingly, we are able to generate these images without providing the model with any direct information about the satellite or radar used to collect imaging data (e.g. antenna size, distance from target, orbital height, signal wavelength, or data sampling rate). In fact, the first component of RDAnet, the deep convolutional encoder, uses standard neural network layers (convolutional layers, pooling layers, and fully connected layers) as well as a well-known architecture, albeit with modified layer activation functions and model regularization approach. Instead, this data is only provided to the model indirectly, through the focused images used to train the model. Additionally, the model did not receive any direct information about the algorithm used to form SAR images used as input to train the model. This suggests that our approach can be applied to other SAR imaging configurations or image formation algorithms, provided that there is sufficient training data.

### 3.2. Impact of Deep Residual Network on image quality

To evaluate the impact of the Deep Residual Network on the quality of images formed by RDAnet, we compared the output RDAnet to two other cases: In the first case, we trained a Deep Convolutional Encoder by itself and applied standard bilinear interpolation to reach the correct image dimensions. For the second case, we used a Deep Residual Network with the same number of layers and filters used in the final RDAnet, but trained it separately from the Deep Convolutional Encoder. Results are shown in Figure 6. Figure 6a shows the SAR image formed using the Range Doppler Algorithm, Figure 6b shows the upsampled output directly from the Deep Convolutional Encoder, Figure 6c shows the output from the Deep Residual Network that was trained separately, while Figure 6d shows the output from RDAnet.

As can be seen by comparing Figure 6b with Figure 6c or Figure 6d, there is clear improvement in image resolution when using the Deep Residual Network single image super resolution approach over standard bilinear interpolation of the output of the Deep Convolutional Encoder. Although the broad structures of the mountain landscape are recovered with the Deep Convolutional Encoder, finer details in the image are blurred together and lost when the Deep Residual Network is not used. Image resolution is further improved when the Deep Convolutional Encoder and Deep Residual Network are trained concurrently. An example of this is seen when the boxed region in Figure 6a is compared with the same region in the other three images. Here, as shown, landscape features are more clearly resolved and the formed image has improved contrast when both components of the complete model are trained together.

To quantitatively assess the impact of the Deep Residual Network, the Structural Similarity (SSIM) index [34] was calculated for each image in the validation set for each of the three cases. Calculation results, which show the average

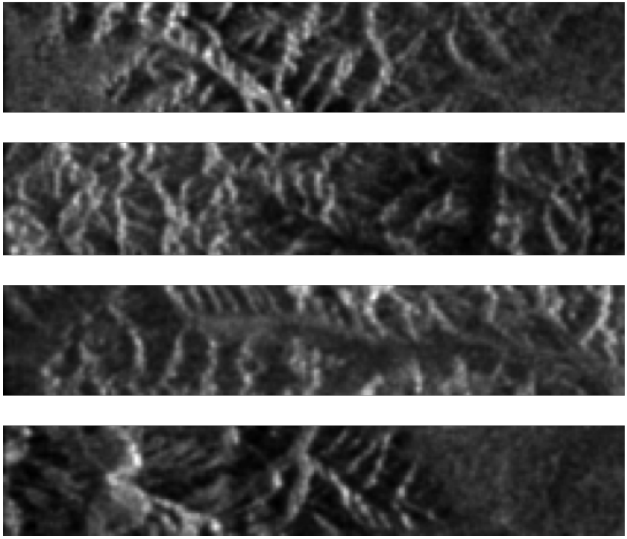 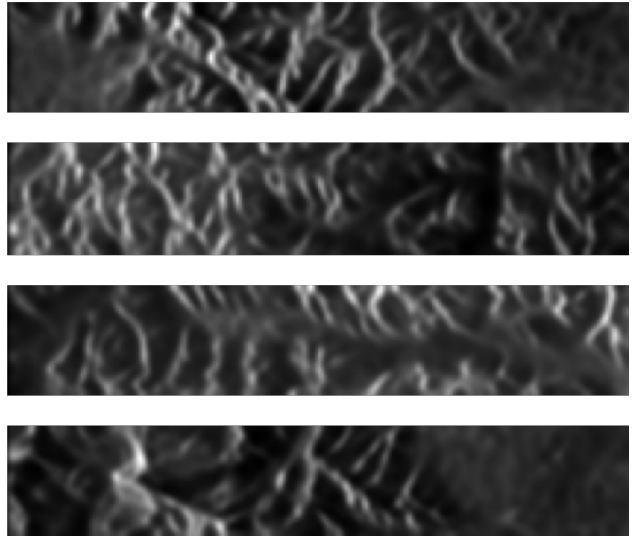

Figure 5: Representative images formed using the Range Doppler Algorithm (left column) and RDAnet (right column). As seen, both images have comparable image quality in terms of resolution and contrast.

SSIM from the three sets of calculations, are shown in Table 1, show that validation set SSIM is highest when the Deep Convolutional Network is trained simultaneously with the Deep Residual Network, which tracks with the qualitative observations discussed above.

Table 1: Quantitative Assessment of impact of Deep Residual Network on formed images

| Interpolation Type | Average SSIM |
|---|---|
| DCE+Bilinear | 0.66 |
| DCE+DRN, trained sequentially | 0.71 |
| DCE+DRN, trained together | 0.85 |

### 3.3. Classification Network Performance

A small hand labeled validation dataset containing 60 mountain landscape images and 43 river landscape images was used to assess the performance of the classification network. After the classification network was trained for 20 epochs, as discussed in Section 2.5, the network was able to classify landscape imagery with an accuracy of 97%.

### 3.4. Computational Evaluation

To evaluate the computational performance of our integrated SAR processing pipeline on an embedded system, RDAnet was integrated with the classification network and complete network forward propagation time was measured on a NVIDIA AGX Jetson Xavier development kit [35].

We found that that timing performance of the model varied depending on the model format. When the model was evaluated using the version of Keras [36] included with TensorFlow, the integrated pipeline operated at a rate of 9 frames per second. However, after we converted the model to TensorFlow's protobuf format, and performed inference with TensorFlow directly, timing performance increased to 23 frames per second. This suggests that our integrated SAR processing pipeline could be suitable for real-time tasks such as change detection or target tracking.

## 4. Discussion

In this work, we demonstrated that SAR images can be formed directly from raw radar data through the use of a deep learning architecture. Unlike [11], which discusses a potential framework for deep learning based SAR image formation but does not contain imaging results from real data, we do not require a forward model of the radar imaging system in the loss function to minimize while training the network. Instead, our network was trained using a standard mean absolute error loss function. This suggests that it should be straightforward to train a RDAnet for any other SAR imaging system with minimal modification to the network design, provided that there is sufficient data for training.

While developing the RDAnet architecture, one challenge we came across was that noise in the raw echo data can cause networks to quickly overfit to the training data. For example, early in the development of RDAnet, we attempted an architecture similar to what is used by AUTOMAP [37] for reconstruction of MRI images from k-space data. This architecture consists of two fully connected layers followed by a sparse convolutional autoencoder. However, we found that regardless of the regularization techniques that we tried, any network trained using an AUTOMAP style architecture did not generalize well when trained with using our echo/image pairs. It is for this reason that we ultimately decided to use a VGG [26] style convolutional encoder for this work. With this architecture, the fully connected layers are below several convolutional and pooling layers, which we believe helps to prevent the model from simply memorizing training data.

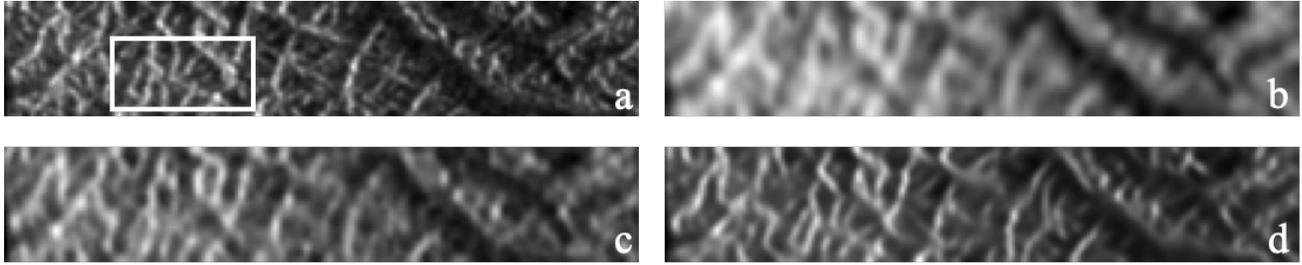

Figure 6: Demonstration of the impact the Deep Residual Network has on formed image quality. a) shows an image formed by the Range Doppler Algorithm, b) shows an image formed with only the Deep Convolutional Encoder, c) shows an image formed by training the Deep Convolutional Encoder and the Deep Residual Network separately, and d) shows an image formed by RDAnet. Image features in the boxed region of 6a are more clearly resolved in 6d than they are in 6b or 6c due to the inclusion and joint training of the deep residual network.

Our integrated SAR pipeline opens the possibility for real-time SAR imaging and processing using a general-purpose computing platform equipped with a GPU. Currently, many solutions for real-time SAR make use of FPGAs [38] or specialized electronics [39] to perform the signal processing needed to form a SAR image. In the computational evaluation of our model, we were able to run our integrated pipeline at 23 frames per second with minimal network optimizations performed. In fact, the only change we made to our model was to convert it from Keras format to a pure TensorFlow model. With additional optimizations, we expect to be able to increase the pipeline's processing rate. For example, the NVIDIA AGX Jetson Xavier development kit provides toolsets to quantize TensorFlow models to 8 bit integer or 16 bit float. We could make use of these tools to both reduce the size of our model and increase pipeline processing rate. As a result, this will enable real-time SAR imaging without the need for specialized hardware.

Furthermore, one interesting aspect of the RDAnet architecture is that the Deep Residual Network component enables a tradeoff between formed image resolution and processing time. In this work, we used 16 residual blocks in our Deep Residual Network. However, the architecture could potentially include fewer blocks, or blocks with fewer filters per block. In either case, we believe this would result in a reduction in image quality in terms of resolution, but also a reduction in time to form the image. Therefore, if the user is interested in real-time or near real-time SAR imaging at a certain frame rate, the number of residual blocks used in the RDAnet can be adjusted in order to meet timing requirements, potentially at the expense of image quality.

Although we were unable to demonstrate this due to lack of appropriate training data, to us, one of the most exciting possibilities opened up by this work is building an integrated SAR pipeline that automatically performs automatic target recognition on a focused SAR image. We envision a processing pipeline where the input is raw echo data and the output is not only a focused SAR image, but an image with accurately classified targets. As mentioned earlier, there is much interest in performing automatic target recognition on SAR images, either through direct classification of targets in SAR images [40], or pixel by pixel semantic segmentation [15]. Currently, much work on the development of SAR automatic target recognition makes use of the MSTAR dataset [6], a collection of SAR images distributed by the Air Force Research Laboratory. This dataset consists of SAR images of 10 different classes of vehicles, collected at different angles. However, images in this dataset have already been formed, and the raw echo data is not publicly available. A similar publicly available dataset containing SAR echo data of different targets would be beneficial to the research community. Likewise, the fact that researchers are still publishing using the MSTAR database [41] despite the fact that it was collected over twenty years ago again highlights the challenge of obtaining SAR data for deep learning research. A publicly available SAR data obtained using a modern SAR imaging system would also be of great value to the research community

## 5. Future Work

During the development of RDAnet, raw SAR echo data was downsampled prior to training to reduce the problem size. Although it enabled us to demonstrate the feasibility of generation of SAR images through a deep learning approach, it also had impact on the resolution of the images reconstructed using the Range Doppler Algorithm that were then used to train our model. Going forward, it would be interesting to train an RDAnet using larger echo/image pairs with less, if any, downsampling to enable formation of even higher resolution SAR images. To maintain the possibility of real-time SAR with larger data size, we will investigate the usage of dilated convolutions [42], which will enable us to increase convolution filter size without increasing the number of calculations that are made.

In addition to improving image resolution, another area we are interested in pursuing is expanding our image formation network to return both magnitude and phase of the focused SAR image. Although magnitude images are sufficient for target classification tasks, there are other SAR applications, such as SAR interferometry [1], that make use of both magnitude and phase of the recovered image.

Another interesting possibility is to investigate the use of GAN [43] style loss functions to train our network. Currently, the RDAnet is trained to minimize mean absolute error. Adding an adversarial discriminator term into the loss function and refining the network by minimizing perceptual loss [44] could help to further improve the quality of images formed using the trained RDAnet.

## 6. Conclusions

In this work, we presented a proof-of-concept demonstration of an integrated SAR processing pipeline, performing both SAR image formation and SAR image processing in a single deep neural network. This pipeline was able to operate in real-time an embedded system without the need for specialized hardware.

Furthermore, as part of this work, we developed RDAnet, a deep learning architecture for the formation of SAR images from raw radar data. After training the RDAnet using real echo data from a mapping satellite we were able to form SAR images of comparable image quality in terms of image resolution and contrast, when compared with images produced using traditional approaches. As far as we know, this is the first demonstration of using a deep learning based approach to form SAR images.

## References

[1] A. Moreira, P. Prats-Iraola, M. Younis, G. Krieger, I. Hajnsek, and K. P. Papathanassiou, "A tutorial on synthetic aperture radar," *IEEE Geoscience and Remote Sensing Magazine*, vol. 1, no. 1, pp. 6–43, Mar. 2013, doi: 10.1109/MGRS.2013.2248301.

[2] R. Bamler and P. Hartl, "Synthetic aperture radar interferometry," *Inverse Problems*, vol. 14, no. 4, pp. R1–R54, Aug. 1998, doi: 10.1088/0266-5611/14/4/001.

[3] V. C. Koo *et al.*, "A NEW UNMANNED AERIAL VEHICLE SYNTHETIC APERTURE RADAR FOR ENVIRONMENTAL MONITORING," *Progress In Electromagnetics Research*, vol. 122, pp. 245–268, 2012, doi: 10.2528/PIER11092604.

[4] H. A. Zebker and R. M. Goldstein, "Topographic mapping from interferometric synthetic aperture radar observations," *Journal of Geophysical Research*, vol. 91, no. B5, p. 4993, 1986, doi: 10.1029/JB091iB05p04993.

[5] E. J. M. Rignot and J. J. van Zyl, "Change detection techniques for ERS-1 SAR data," *IEEE Transactions on Geoscience and Remote Sensing*, vol. 31, no. 4, pp. 896–906, Jul. 1993, doi: 10.1109/36.239913.

[6] T. D. Ross, S. W. Worrell, V. J. Velten, J. C. Mossing, and M. L. Bryant, "Standard SAR ATR evaluation experiments using the MSTAR public release data set," presented at the Aerospace/Defense Sensing and Controls, Orlando, FL, 1998, pp. 566–573, doi: 10.1117/12.321859.

[7] H. Espedal, "Detection of oil spill and natural film in the marine environment by spaceborne SAR," in *IEEE 1999 International Geoscience and Remote Sensing Symposium. IGARSS'99 (Cat. No.99CH36293)*, Hamburg, Germany, 1999, vol. 3, pp. 1478–1480, doi: 10.1109/IGARSS.1999.771993.

[8] L. A. Gorham and L. J. Moore, "SAR image formation toolbox for MATLAB," presented at the SPIE Defense, Security, and Sensing, Orlando, Florida, 2010, p. 769906, doi: 10.1117/12.855375.

[9] S.-J. Wei, X.-L. Zhang, J. Shi, and G. Xiang, "SPARSE RECONSTRUCTION FOR SAR IMAGING BASED ON COMPRESSED SENSING," *Progress In Electromagnetics Research*, vol. 109, pp. 63–81, 2010, doi: 10.2528/PIER10080805.

[10] "Range Doppler Algorithm." [Online]. Available: https://earth.esa.int/handbooks/asar/CNTR2-6-1-2-3.html.

[11] E. Mason, B. Yonel, and B. Yazici, "Deep learning for radar," in *2017 IEEE Radar Conference (RadarConf)*, Seattle, WA, USA, 2017, pp. 1703–1708, doi: 10.1109/RADAR.2017.7944481.

[12] B. Yonel, E. Mason, and B. Yazici, "Deep Learning for Passive Synthetic Aperture Radar," *IEEE J. Sel. Top. Signal Process.*, vol. 12, no. 1, pp. 90–103, Feb. 2018, doi: 10.1109/JSTSP.2017.2784181.

[13] S. Chen, H. Wang, F. Xu, and Y.-Q. Jin, "Target Classification Using the Deep Convolutional Networks for SAR Images," *IEEE Trans. Geosci. Remote Sensing*, vol. 54, no. 8, pp. 4806–4817, Aug. 2016, doi: 10.1109/TGRS.2016.2551720.

[14] H. Furukawa, "Deep Learning for End-to-End Automatic Target Recognition from Synthetic Aperture Radar Imagery," *arXiv:1801.08558 [cs]*, Jan. 2018.

[15] H. Furukawa, "SAVERS: SAR ATR with Verification Support Based on Convolutional Neural Network," *arXiv:1805.06298 [cs]*, May 2018.

[16] "TensorFlow." [Online]. Available: https://www.tensorflow.org/.

[17] "PyTorch." [Online]. Available: https://pytorch.org/.

[18] "Cloud Tensor Processing Units." [Online]. Available: https://cloud.google.com/tpu/docs/tpus.

[19] A. Moreira, "Improved multilook techniques applied to SAR and SCANSAR imagery," *IEEE Transactions on geoscience and remote sensing*, vol. 29, no. 4, pp. 529–534, 1991.

[20] D. Glasner, S. Bagon, and M. Irani, "Super-resolution from a single image," in *2009 IEEE 12th International Conference on Computer Vision*, Kyoto, 2009, pp. 349–356, doi: 10.1109/ICCV.2009.5459271.

[21] "Alaska Satellite Facility." [Online]. Available: https://www.asf.alaska.edu.

[22] "ERS Mission Summary." [Online]. Available: https://earth.esa.int/web/guest/missions/esa-eo-missions/ers/mission-summary.

[23] R. Birk, W. Camus, E. Valenti, and W. McCandless, "Synthetic aperture radar imaging systems," *IEEE Aerospace and Electronic Systems Magazine*, vol. 10, no. 11, pp. 15–23, Nov. 1995, doi: 10.1109/62.473408.

[24] "ASF Vertex Tool." [Online]. Available: https://search.asf.alaska.edu/.

[25] "MATLAB resample function." [Online]. Available: https://www.mathworks.com/help/signal/ref/resample.html.

[26] K. Simonyan and A. Zisserman, "Very Deep Convolutional Networks for Large-Scale Image Recognition," *arXiv:1409.1556 [cs]*, Apr. 2015.

[27] K. He, X. Zhang, S. Ren, and J. Sun, "Delving Deep into Rectifiers: Surpassing Human-Level Performance on ImageNet Classification," *arXiv:1502.01852 [cs]*, Feb. 2015.

[28] J. Tompson, R. Goroshin, A. Jain, Y. LeCun, and C. Bregler, "Efficient Object Localization Using Convolutional Networks," *arXiv:1411.4280 [cs]*, Jun. 2015.

[29] N. Srivastava, G. Hinton, A. Krizhevsky, I. Sutskever, and R. Salakhutdinov, "Dropout: a simple way to prevent neural networks from overfitting," *The journal of machine learning research*, vol. 15, no. 1, pp. 1929–1958, 2014.

[30] B. Lim, S. Son, H. Kim, S. Nah, and K. M. Lee, "Enhanced Deep Residual Networks for Single Image Super-Resolution," *arXiv:1707.02921 [cs]*, Jul. 2017.

[31] K. He, X. Zhang, S. Ren, and J. Sun, "Deep Residual Learning for Image Recognition," *arXiv:1512.03385 [cs]*, Dec. 2015.

[32] W. Shi *et al.*, "Real-time single image and video super-resolution using an efficient sub-pixel convolutional neural network," in *Proceedings of the IEEE conference on computer vision and pattern recognition*, 2016, pp. 1874–1883.

[33] D. P. Kingma and J. Ba, "Adam: A Method for Stochastic Optimization," *arXiv:1412.6980 [cs]*, Dec. 2014.

[34] Z. Wang, A. C. Bovik, H. R. Sheikh, and E. P. Simoncelli, "Image Quality Assessment: From Error Visibility to Structural Similarity," *IEEE Trans. on Image Process.*, vol. 13, no. 4, pp. 600–612, Apr. 2004, doi: 10.1109/TIP.2003.819861.

[35] "NVIDIA AGX Xavier." [Online]. Available: https://developer.nvidia.com/embedded/jetson-agx-xavier-developer-kit.

[36] F. Chollet and others, *Keras*. 2015.

[37] B. Zhu, J. Z. Liu, S. F. Cauley, B. R. Rosen, and M. S. Rosen, "Image reconstruction by domain-transform manifold learning," *Nature*, vol. 555, no. 7697, pp. 487–492, Mar. 2018, doi: 10.1038/nature25988.

[38] Xin Xiao, Rui Zhang, Xiaobo Yang, and Gang Zhang, "Realization of SAR real-time processor by FPGA," in *IEEE International IEEE International IEEE International Geoscience and Remote Sensing Symposium, 2004. IGARSS '04. Proceedings. 2004*, Anchorage, AK, USA, 2004, vol. 6, pp. 3942–3944, doi: 10.1109/IGARSS.2004.1369989.

[39] B. Walker, G. Sander, M. Thompson, B. Burns, R. Fellerhoff, and D. Dubbert, "A high-resolution, four-band SAR Testbed with real-time image formation," in *IGARSS '96. 1996 International Geoscience and Remote Sensing Symposium*, Lincoln, NE, USA, 1996, vol. 3, pp. 1881–1885, doi: 10.1109/IGARSS.1996.516827.

[40] D. A. E. Morgan, "Deep convolutional neural networks for ATR from SAR imagery," presented at the SPIE Defense + Security, Baltimore, Maryland, United States, 2015, p. 94750F, doi: 10.1117/12.2176558.

[41] L. Zou and Z. Qiao, "Convergence analysis of the CNN algorithm in target recognition using SAR images," in *Radar Sensor Technology XXIII*, Baltimore, United States, 2019, p. 32, doi: 10.1117/12.2519086.

[42] P. Wang *et al.*, "Understanding Convolution for Semantic Segmentation," in *2018 IEEE Winter Conference on Applications of Computer Vision (WACV)*, Lake Tahoe, NV, 2018, pp. 1451–1460, doi: 10.1109/WACV.2018.00163.

[43] I. Goodfellow *et al.*, "Generative Adversarial Nets," in *Advances in Neural Information Processing Systems 27*, Z. Ghahramani, M. Welling, C. Cortes, N. D. Lawrence, and K. Q. Weinberger, Eds. Curran Associates, Inc., 2014, pp. 2672–2680.

[44] J. Johnson, A. Alahi, and L. Fei-Fei, "Perceptual losses for real-time style transfer and super-resolution," in *European conference on computer vision*, 2016, pp. 694–711.